\documentclass[useAMS,usenatbib,fleqn]{mn2e}
\usepackage{amsmath,amssymb,amsfonts,latexsym}
\usepackage{graphicx}
\usepackage{longtable}
\usepackage{fixltx2e}
\usepackage{natbib}
\usepackage[draft]{hyperref}

\def\apjs{{The Astrophysical Journal Supplement}}
\def\apjl{{ApJL}}

\def\aap{{A\&A}}

\def\04a{{2004 a}}
\def\04b{{2004 b}}

\title[Magnetic spots at the neutron star surface]{Creation of magnetic spots at the neutron star surface}

\author[U.~Geppert \& D.~Vigan\`{o}]{U.~Geppert$^{1,2}$\thanks{E-mail:ulrich.geppert@dlr.de}, D.~Vigan\`{o}$^{3}$\\
$^1$ Kepler Institute of Astronomy, University of Zielona G\'{o}ra, Lubuska 2, 65-265, Zielona G\'{o}ra, Poland \\
$^2$ German Aerospace Center, Institute for Space Systems, Robert-Hooke-Str. 7, 28359 Bremen, Germany\\
$^3$ Institute of Space Sciences (CSIC--IEEC), Campus UAB, Faculty of Science, Torre C5-parell, 08193 Bellaterra, Spain}

\begin{document}

\date{}

\maketitle

\label{firstpage}
\begin{abstract}
According to the partially screened gap scenario, an efficient electron-positron pair creation, a general precondition of radio-pulsar activity, relies on the existence of magnetic spots, i.e., local concentrations of strong and small scale magnetic field structures at the surface of neutron stars. They have a strong impact on the surface temperature, which is potentially observable. Here we reinforce the idea that such magnetic spots can be formed by extracting magnetic energy from the toroidal field that resides in deep crustal layers, via Hall drift. We study and discuss the magneto-thermal evolution of qualitatively different neutron star models and initial magnetic field configurations that lead to the creation of magnetic spots. We find that magnetic spots can be created on a timescale of $10^4$ years with magnetic field strengths $\gtrsim 5\times 10^{13}$ G, provided almost the whole magnetic energy is stored in its toroidal component, and that the conductivity in the inner crust is not too large. The lifetime of the magnetic spots is at least $\sim$one million of years, being longer if the initial field permeates both core and crust.
\end{abstract}

\begin{keywords}
stars: pulsars: general -- stars: magnetic field
\end{keywords}

\section{Introduction}
Most of the models explaining the non-thermal emission of neutron stars (NSs) in radio, $X$-ray and $\gamma$-ray rely on the presence of an inner acceleration gap, where an efficient pair production and a strong electric acceleration are indispensable prerequisites (see e.g. \citealt{ML10} and references therein). In this scenario, the magnetic field plays an important role, since it partly determines the intensity of the electric force.
In young and/or highly magnetized NSs the large-scale dipolar magnetic field (inferred from timing properties) is strong enough to support the gap. However, the large curvature radius of a purely dipolar magnetic field impedes the creation of a sufficiently large number of $e^-$-$e^+$ pairs. Therefore, a self-consistent gap model requires a local, strong magnetic field close to the surface, in form of small-scale, high multipoles and/or toroidal components, which do not affect the torque on the star, i.e., its observable spin-down. Such small-scale fields are an essential ingredient of the two basically different radio pulsar models: the space charge limited flow (SCLF) model \citep{AS79} and the vacuum gap model \citep{RS75} with its refinement, the partially screened gap (PSG) model \citep{GMG03}. The latter takes into account the $B$-dependence of the cohesive energy that binds the charges at the polar cap surface, while the SCLF model assumes zero cohesive energy there. \cite{CB14} has recently performed particle-in-cell simulations of the whole magnetosphere, proving that, if particles can be extracted freely from the surface, then no polar-cap gaps are formed. \cite{GMG03} showed that polar cap gaps can be formed even if the density is only slightly below the Goldreich-Julian value (due to the non-negligible cohesive energy of ions and a relatively low temperature), and it is compatible with the drifting sub-pulse phenomena observed in radio pulsars. The condensed surface requirement in the PSG model implies an additional constraint on the local intensity of the small scale magnetic field strength.

The typical parameters required in the magnetic spots are: strength $B_s\gtrsim 5\times 10^{13}$ G (for $T_s\gtrsim 10^6$ K) \citep{ML07} and a curvature radius of the field lines $R_{\text{curv}} \lesssim 10$ km \citep{GMG03}, which is much smaller then the value for a dipole close to the magnetic pole, $R_{\text{curv}} \gtrsim 100$ km. Both the requirements for such spots and a possible physical mechanism which could create them, the crustal Hall drift of the magnetic field, have been discussed in detail by \citet{GGM13}. The Hall drift displaces the crustal currents, and allow the interchange of magnetic energy between poloidal and toroidal components (see, e.g., \citealt{RBPAL07,VRPPAM13}). Since the thermal and magnetic evolutions in the crust are tightly connected to each other, the creation of magnetic spots has also consequences for the potentially in X-rays observable surface temperature distribution.

X-ray observations of isolated NSs reveal that their surfaces are not isothermal (see e.g. \cite{YP04}, \citealt{VPRP14} and references therein). For several sources, like the three musketeers \citep{deluca05}, their X-ray spectra can be satisfactory fitted by models that include two blackbodies, where the hotter temperature is attributed to a significantly smaller portion of the surface than the lower one. Other hints, like large pulsed fractions of the thermal radiation seen in many observations and small emitting radii (often of the order of $\sim 0.1-$few km, as inferred from the X-ray spectral fitting, to be compared with the expected NS radius $\sim 10-13$ km), are frequent in pulsars, regardless whether they are highly or weakly magnetized, young or old, radio-loud or radio-quiet\footnote{Indeed, pulsed fractions of thermally emitting NSs may vary in a wide range of $1\ldots 20 \%$ \citep{K08}.}.

One reason for an inhomogeneous distribution of the surface temperature, $T_s$, can be understood by studying the heat transport in the outer layers of a NS (crust and envelope): it becomes highly anisotropic in the presence of a sufficiently strong magnetic field ($B$), strongly reducing the conductivity in the direction perpendicular to its lines. Thus, in order to create observable differences in the $T_s$-distributions, the crustal magnetic field needs to have large scale components tangential to the surface, e.g., strong dipolar toroidal components \citep{GKP06,PMP06}. This scenario is especially suitable for highly magnetized NSs (magnetars, \citealt{M08}; high-B pulsars, \citealt{NK11}; X-ray isolated NSs (XINSs), \citealt{T09}).

Spectra with very small inferred emitting radii ($R_{bb}\lesssim 1$ km), as seen in several old radio-pulsars \citep{ZP04b}, can be explained by the heat deposited by the back-flow of ultra-relativistic magnetospheric particles which are accelerated in the potential drop above the pole \citep{GMZ07b}. The bombarding particles are electrons or positrons which are produced in pair creation processes by the interaction of $\gamma$-ray photons with the strong and sufficiently curved magnetic field just above the polar cap \citep{S71,S13}.

However, it is not clear whether the formation of such spots can happen only under very specific assumptions on the poorly known initial magnetic field configuration and/or NS structure. The aim of the paper is to study further the initial conditions for which the magneto-thermal evolution leads to the creation of long-living magnetic spots. For this purpose, we present some results obtained with the 2D magneto-thermal evolution code described in \citet{VPM12,VRPPAM13}.

In order to 'switch-on' a radio pulsar, the large scale open field lines have to be connected to the strong and small scale field of magnetic spots. Only in such a case the inner acceleration region can be formed which consequently leads to production of plasma responsible for radio emission. Whether the generated plasma is suitable to produce radio emission is beyond the scope of the present study.

The plan of the paper is as follows: in \S~\ref{sec:mt} we briefly review the basic equations of the magneto-thermal evolution, discussing the initial magnetic field configuration and the microphysical setup. The results of the simulations are presented in \S~\ref{sec:results}. Conclusions and open issues are drawn in \S~\ref{sec:conclusions}.

\section{Magneto-thermal evolution of the crust}\label{sec:mt}

\subsection{Basic equations}
In the crust the magnetic field $\vec B$ is tied to the electrons which circulate in currents through a crystalline lattice
formed by almost immobile ions. Therefore, the only processes that drive the magnetic evolution are Ohmic 
diffusion/dissipation and Hall drift (see e.g. \citealt{GR92,RG02,HR02,PG07}), described by the Hall induction equation:

\begin{equation}
\frac{\partial\vec B}{\partial t}= - \vec{\nabla} \times\left[\frac{c^2}{4\pi\sigma}\vec{\nabla} \times (e^{\nu}\vec{B}) + 
\frac{c}{4\pi e n_e} [(\vec{\nabla} \times (e^{\nu}\vec{B})] \times \vec{B}  \right],
\label{Hallind}
\end{equation}
where $\sigma$ denotes the electric conductivity, dependent on the local temperature $T$, the density $\rho$, and the composition of the crustal ionic lattice. The second term in the r.h.s. of equation~(\ref{Hallind}) represents the Hall drift, whose pre-factor only depends on the crustal electron number density profile $n_e$. The gravitational redshift factor $e^{\nu}$ is given by the structure of the star. All coefficients in equation~(\ref{Hallind}) are strongly varying functions of the radial coordinate $r$. The electrical current is given by $\vec{J}=ce^{-\nu}\vec{\nabla}\times (e^{\nu}\vec{B})/4\pi$.

The mutual dependence of thermal and magnetic evolution is also seen in the thermal balance equation that
describes the evolution of the crustal temperature $T$

\begin{equation}
c_v e^{\nu}\frac{\partial T}{\partial t} - \vec{\nabla} \cdot \left[e^{\nu}\hat{\kappa}\cdot \vec{\nabla}(e^{\nu}T)\right]
=e^{2\nu}\left(-Q_{\nu}+Q_h\right)~,
\label{Tevol}
\end{equation}
where $c_v$ is the specific heat. The coupling to the magnetic evolution is given by the $\vec{B}$-dependent components of the heat conductivity tensor $\hat\kappa$ \citep{GKP04}, the Joule heating, $Q_h=J^2/\sigma$ \citep{PMG09,VRPPAM13}, and, to a lesser extent, by weak $B$-dependences of some of the processes contributing to the neutrino luminosity $Q_{\nu}$.

In order to solve the coupled non-linear partial differential equations (\ref{Hallind}) and (\ref{Tevol}), we use the numerical 2D code by \cite{VPM12}. Up to now no 3D code version is available, therefore all calculations are performed in axial symmetry, i.e. the magnetic/hot spots are indeed polar caps or rings at other latitudes. All details regarding the numerics and boundary conditions can be found in \citet{VPM12}, while the micro-physics setup (equation of state, electric and thermal conductivity, thermoelectric coefficients, specific heat, envelope model, neutrino emissivities, onset of superfluidity) is  extensively described in \cite{APM08b}, \cite{VRPPAM13}, and \cite{V13}. Some microphysical inputs ($c_V$ and conductivities) are provided by the publicly available routines developed by the Potekhin group \citep{PC13}.\footnote{See \url{http://www.ioffe.ru/astro/EIP/index.html} and \url{http://www.ioffe.ru/astro/conduct}}

The main contributions to the thermal and electric conductivities inside the NS are two \citep{HUY90,SYHP07,HCB09}. The first one is the electron-phonon scattering, whose rate is highly dependent on the temperature, so that, when the latter is high enough, this process dominates the conductivity. On the other hand, temperature-independent collisions between electrons and crystal impurities, dislocations, and other irregularities can be important for low temperatures and/or for highly impure lattice. The content of impurity is quantified by the parameter $Q$, which is theoretically largely uncertain, because it depends on the poorly known geometry and composition of the ionic lattice.

Actually, the value of $Q$ can be indirectly constrained by observations. Fitting the cooling light curve of a few low mass X-ray binaries after long accretion stages, \cite{BC09} and \cite{TAP13} estimate an impurity parameter $Q\sim 3-5$ in the inner crust. On the other side, \cite{PVR13} suggested that the innermost part of the crust, close to the interface with the core, has to be highly resistive, in order to explain the rapid magnetic field decay responsible for a clustering of periods below $\sim 12$~s in isolated X-ray pulsars. The magneto-thermal simulations are in agreement with the original idea of such fast decay of magnetic field \citep{CGP00} only for large values of $Q_{\rm ic} \sim 10-100$ in the innermost region of the crust, where the bulk of electric current circulates due to the Hall drift \citep{VRPPAM13}. Such large values are compatible with an amorphous structure of the crustal lattice \citep{J88}, or the presence of pasta phase (see the discussion in \citealt{PVR13}).

Thus, similarly to what done by \cite{PVR13} and \cite{VRPPAM13}, we assume in the outer crust a low value $Q_{\rm oc}=5$ for $\rho < 5\times 10^{13}$g cm$^{-3}$, and $Q_{\rm ic}=25$ in the inner crust layers $\rho > 5\times 10^{13}$g cm$^{-3}$.

\subsection{The initial magnetic field configuration.}\label{sec:initial_b}

What is the most realistic choice for the initial magnetic field configurations, from which the magneto-thermal evolution should begin? The answer could come from theoretical (analytical and numerical) studies of the initial magnetic field generation and dynamics, which is very challenging. Theoretically, the creation of strong fields (especially toroidal) should be possible during the first $\sim 20$ seconds of a NS's life, when convective motion and differential rotation is present and able to wind up an initially relatively weak poloidal field component to an extremely strong toroidal one \citep{P79,S02,OJA14}. The rapid rotation soon after birth can help to stabilize a dominant, very strong toroidal field component against Tayler instabilities \citep{B06}.

Numerical simulations of the first instant after the SN explosion are computationally very expensive. The most recent and longest MHD simulations of supernova explosion, including the amplification of magnetic field, follow the proto-NS evolution during almost one second \citep{OJA14}. These simulations show a very complicated magnetic field configuration, far from being dipolar, which, at least during this phase, does not approach a steady state. 

After the initial proto-NS phase, convective motions of matter almost cease (except, probably, in the envelope), the whole star is liquid. During this phase, the Alfv\'{e}n time is short, so that a MHD equilibrium is thought to be established before the temperature drop leads to the crystallization of the crust and/or to transitions to superfluid/superconductive phases.

The magnetic configuration of axisymmetric MHD equilibria has been discussed, e.g., by \cite{L10,GAL12,LJ12,FYE12,CR13}. A common feature of such equilibria is that the toroidal field has to be confined within the poloidal field lines which close inside the star, in order to match with a potential magnetic field outside.

\citet{BN06} argue that such twisted torus configuration is stable only if the poloidal and toroidal component have locally about the same field strengths. However, these configurations (for some of which the stability has not been proven), are likely to be not unique, and it is not clear towards which one the star will go after birth. They also rely on a number of assumptions and/or simplification (no superconductivity). For example, qualitatively different configurations can be found by relaxing the assumption of no currents circulating in the magnetosphere (i.e., prescribing a non-potential magnetic field outside the star, \citealt{GLA14}).

Recently, \cite{GCRALV13} and \cite{GC14a} performed detailed studies on MHD and Hall equilibria that could be established in NS crusts, their interplay and  how their evolution depends on the choice of the initial conditions. \cite{GCRALV13} have shown that, even if at the beginning of the Hall-MHD evolution the field is in MHD equilibrium, the presence of a strong radial gradient in the electron fraction will unavoidably create out of dipolar ones higher order toroidal field modes.

\cite{L14} has recently shown that a more realistic treatment of the interface between the crust and the type-II superconducting core can lead to equilibria where a larger fraction of the magnetic field is in the crust, especially for lower values of the external dipole. This is an example of the influence of the poorly understood effects which likely regulate the magnetic configuration and evolution in the core. In the absence of any clear prescription, we naively assume that the magnetic field in the core, if present, evolves only due to the Ohmic dissipation, which is so slow there, that the magnetic field of the core is basically frozen during timescales of tens of Myr at least.

Even if we knew what is the actual MHD equilibrium, there are more complications. As a matter of fact, quite soon the outer layers of the star begin to form a crust by crystallization. Whether the magnetic field is frozen close to the MHD equilibrium structure during the process of solidification is unclear. The latter starts at about nuclear density, i.e. at the future crust-core interface, after minutes from birth \citep{APM08b}, while it can take months or years for the outer layers. The expanding crystalline shell, in which the field evolution is determined by the Hall-MHD, is confined from below and above by liquid material where the field has supposedly reached an MHD equilibrium. The situation becomes even more complicated after superfluid/superconducting phase transitions appear. 

\begin{table*}
\caption{Parameter of NS models, differing in the NS mass and in the initial magnetic field configuration, shown in Fig.~\ref{fig:initial_b}. $B_{\rm dip}$ denotes the strength of the dipolar poloidal field component at the polar surface, while max$(B_{\rm pol}^{\rm cr})$, max$(B_{\rm tor})$ and max$(B_{\rm tor}^{\rm cr})$ are the maximum strengths of the poloidal component in the crust, of the toroidal component in the whole star and in the crust only, respectively. We also indicate the ratio of the magnetic energy stored in the toroidal component to the total magnetic energy (considering both core and crust).}
\begin{tabular}{lccccccc} 
\hline 
\hline\noalign{\smallskip} 
Model & $M$ & {configuration} & $B_{\rm dip}$ & max($B_{\rm tor}$) & max($B_{\rm pol}^{\rm cr}$) & max($B_{\rm tor}^{\rm cr}$) & $E_{\rm tor}/E_B$  \\ 
 & ($M_\odot$) &  & ($10^{13}$ G) & ($10^{13}$ G) & ($10^{13}$ G) & ($10^{13}$ G) & $\%$ \\
\hline 
AL & 1.40 & crust confined  & 1 & 150 & 11  & 150 & 99.386 \\ 
QL & 1.40 & crust confined & 1 & 150 & 11 & 150 & 99.284\\
BL & 1.40 & core permeating  & 1 & 2000 & 1.2 & 1490 & 99.998\\
BH & 1.76 & core permeating & 1 & 2000 & 1.1 & 820 & 99.998\\
\hline
\hline\noalign{\smallskip} 
\end{tabular} 
\label{tab:models} 
\end{table*} 

\begin{figure*}
\centering
\includegraphics[width=7.cm]{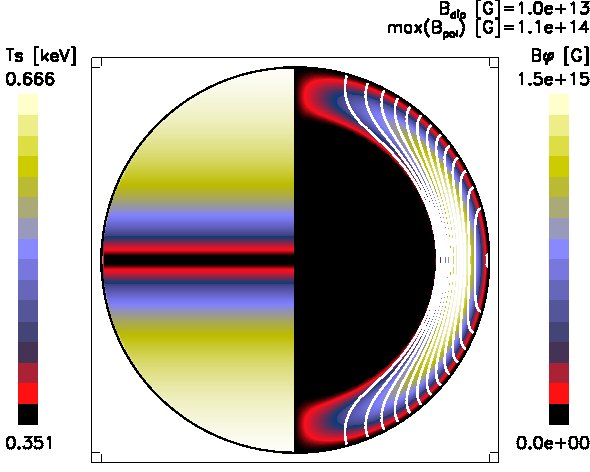}
\includegraphics[width=7.cm]{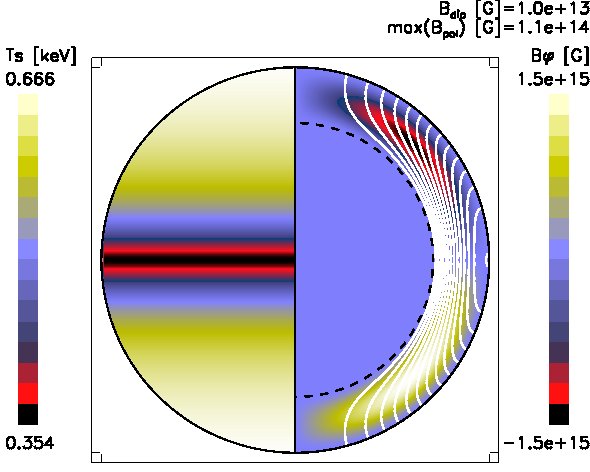}\\
\includegraphics[width=7.cm]{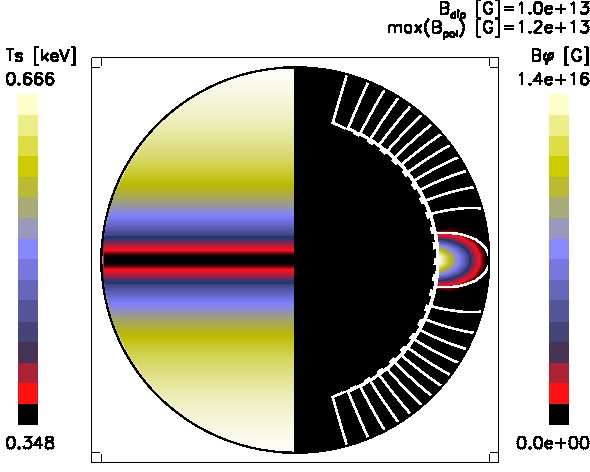}
\includegraphics[width=7.cm]{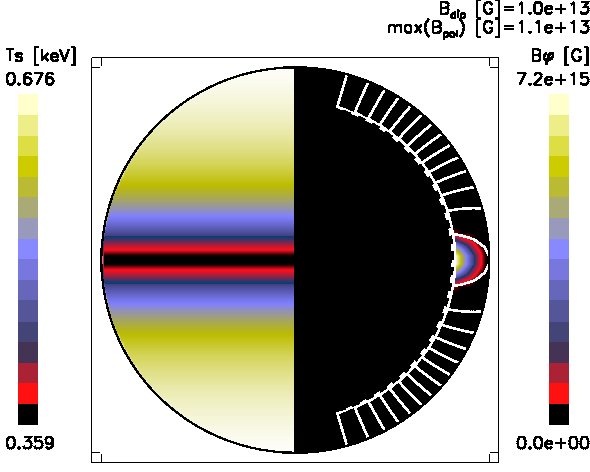}
\caption{Initial magnetic field configuration for crust-confined models AL (top left), QL (top right), and for the core-threading models BL (bottom left) and BH (bottom right). The left hemisphere represents the surface temperature, while the right hemisphere show the crustal magnetic field configuration. For the latter, we show only the crustal region, which is stretched by a factor of 4. In models BL and BH, lines penetrate in the core (not shown). The poloidal magnetic field lines are drawn in white solid lines, while colors maps the intensity of the toroidal magnetic field $B_\varphi$.}
\label{fig:initial_b}
\end{figure*}

All the points listed above raise a question whether the classical twisted torus equilibrium configurations are realistic. Given the highly theoretical uncertainties, one can instead rely on observational hints. In this sense, a basic property needed to explain a variety of NS observed phenomenology is the concentration of a huge amount of electric current in the crust. In the latter, the Ohmic decay timescale is orders of magnitude shorter than in the core. This allows to understand, at the same time, the large luminosity observed in highly magnetized NSs (high-B pulsars and magnetars), the distribution of $P$ and $\dot{P}$ for isolated NSs \citep{VRPPAM13} and the outburst activity caused by the fractures \citep{PP11} or plastic deformations \citep{BL14} of the magnetically stressed crust.

Moreover, dominant, large-scale toroidal components in the crust, larger than the poloidal ones by orders of magnitude, have been proposed to explain the large thermal pulsed fraction of the NS in the Supernova Remnant KES 79 \citep{SL12}, and the magnetar-like outburst activity of the apparently weakly magnetized NS SGR~0418+5729 \citep{RETIZSMTGGK10}. The study of braking indices of radio pulsars at different ages  yield another indication of a toroidal field spread throughout the crust \citep{PVG12,GC14b}. All these observed properties point toward a concentration of the currents in the crust throughout the life of a NS. Such configuration could be supplied by either some (poorly understood) mechanisms of flux expulsion form the superconducting core, or by strongly multipolar and twisted initial geometry, at least in the crust.

\subsection{Neutron star models}

We consider four NS models which differ in mass and initial magnetic field configurations. For all models, we start with an initial, isotropic temperature $T_{\rm in}=10^{10}$ K in the interior (note that the strong neutrino losses make the evolution completely insensitive on the specific value chosen for the initial temperature, in the range of expected proto-NS temperature, $T_{\rm in}\sim 10^{10}-5\times 10^{11}$ K, \citealt{PRPLM99}). We use an iron envelope model, fixing its interface with the crust at $\rho=3\times 10^{10}$ g cm$^{-3}$. The grid where we map the magnetic field consists of 140 angular $\times$ about 100 radial points. The magneto-thermal evolution of the models is tracked for a few millions years, a typical age of radio pulsars. 

The characterizing parameters are shown in Table~\ref{tab:models}. The first important difference between the models is the initial magnetic field geometry. Given the virtually total uncertainty about the initial magnetic field configuration (see \S~\ref{sec:initial_b}), we explore two qualitatively different, simplified geometries. They are presented in Fig.~\ref{fig:initial_b}. 

In models AL (top left panel) and QL (top right panel), the magnetic field is completely confined to the crust, with differences in the polarity of the toroidal field: dipolar vs quadrupolar, respectively. Models BL (bottom left panel) and BH (bottom right panel), instead, start with a twisted torus configuration, i.e., a magnetic field that penetrates into the core and its toroidal component is located only within closed poloidal field lines, qualitatively mimicking the conditions of the most popular axisymmetric MHD equilibrium. Note that, in models BL and BH, magnetic field lines are more stretched and the currents are mainly spread in the highly conductive core. In all models, initial poloidal and toroidal fields are purely dipolar, and more than $99\%$ of the total magnetic energy is stored in the toroidal field.

For all models, we fix the dipolar, poloidal component of the magnetic field (responsible for the spin-down of the NS) to be $B_{\rm dip}=10^{13}$ G at the polar surface. In models AL and QL, the maximum of the toroidal field strength is max$(B_{\rm tor})=1.5\times 10^{15}$ G, located in the inner crust, at $\rho \approx 10^{13}$g cm$^{-3}$. In models BH and BL, we fix max($B_{\rm tor})=2\times 10^{16}$ G, and such maximum intensity lies in the outer core, at $\rho \approx 3\times 10^{14}$g cm$^{-3}$. In Table~\ref{tab:models} we also show the maximum value of the toroidal and poloidal field in the crust, max($B_{\rm tor}^{\rm cr}$) and max($B_{\rm pol}^{\rm cr}$), respectively, and the ratio of magnetic energy stored in the toroidal component, which is close to 1.

\begin{figure*}
\centering
\includegraphics[width=5.cm]{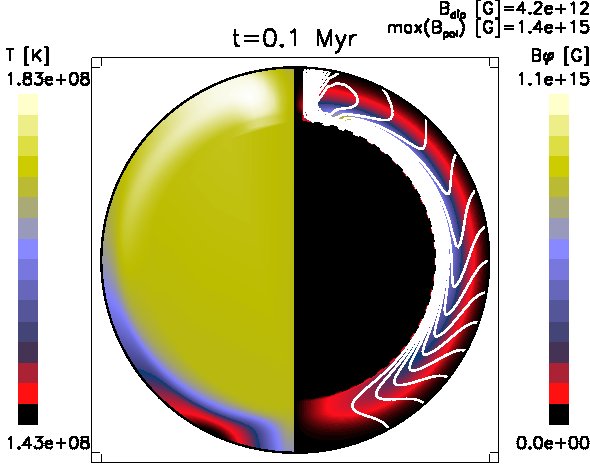}
\includegraphics[width=5.cm]{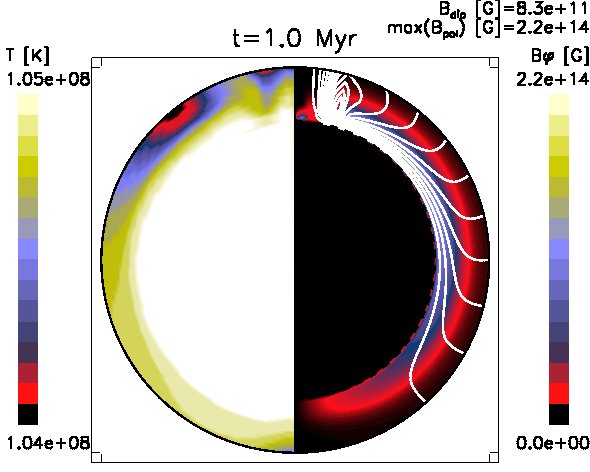}
\includegraphics[width=5.cm]{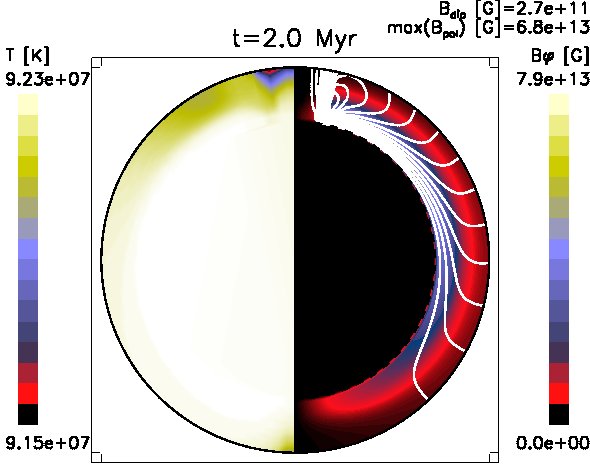}\\
\vspace{10mm}
\includegraphics[width=5.cm]{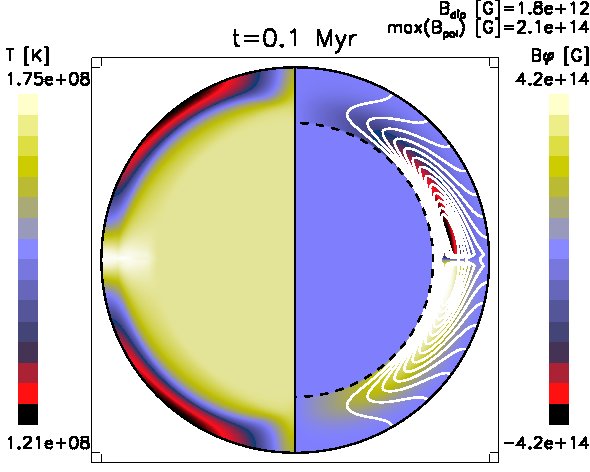}
\includegraphics[width=5.cm]{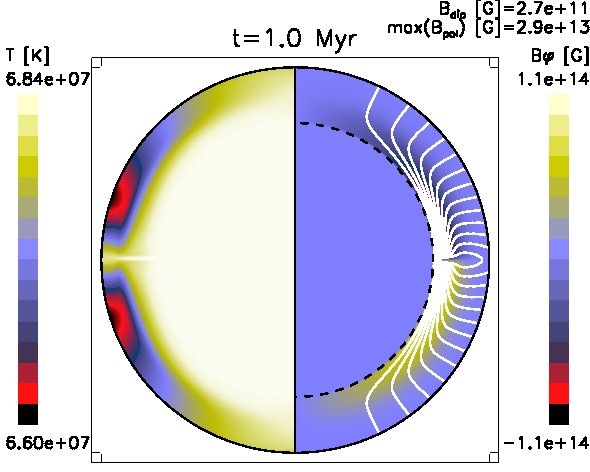}
\includegraphics[width=5.cm]{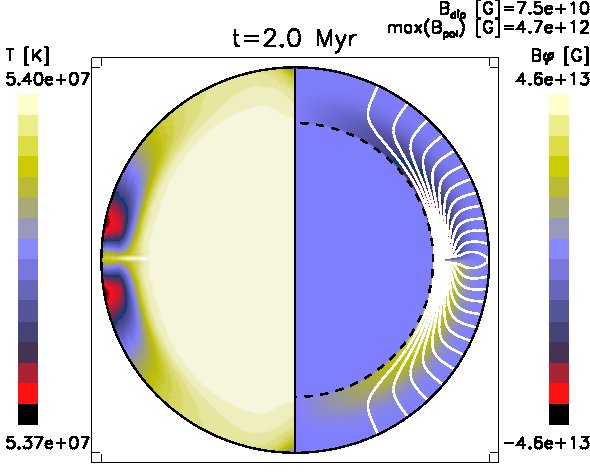}\\
\vspace{10mm}
\includegraphics[width=5.cm]{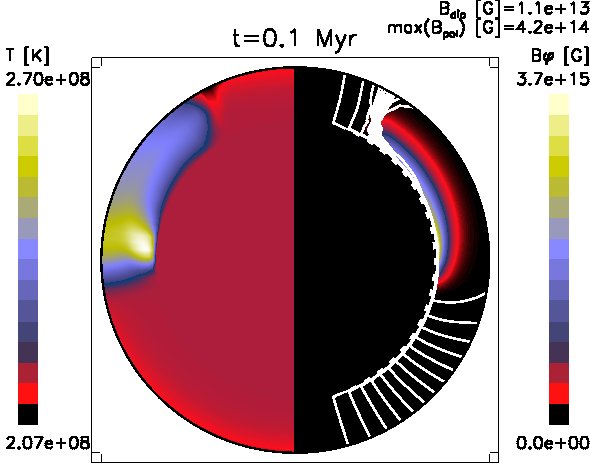}
\includegraphics[width=5.cm]{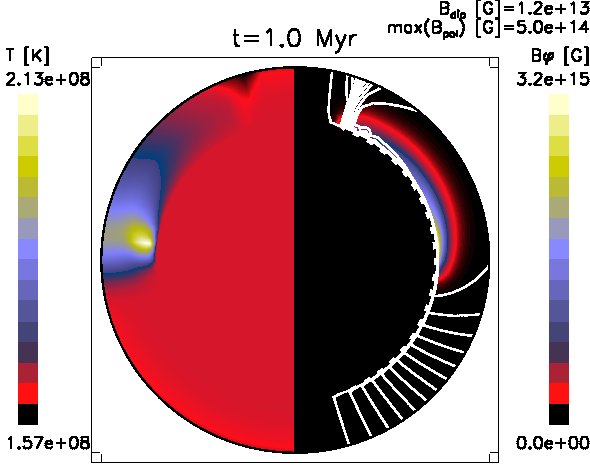}
\includegraphics[width=5.cm]{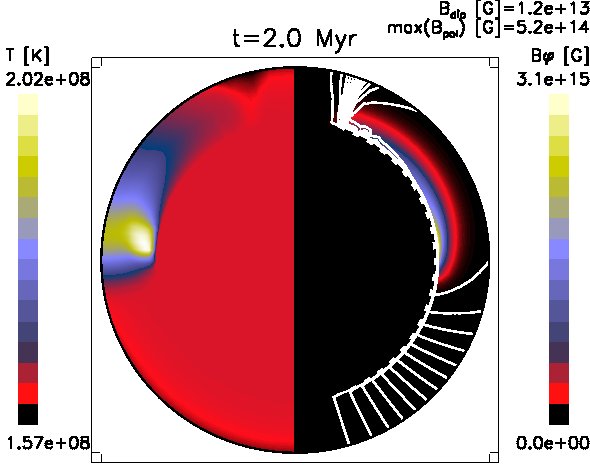}\\
\vspace{10mm}
\includegraphics[width=5.cm]{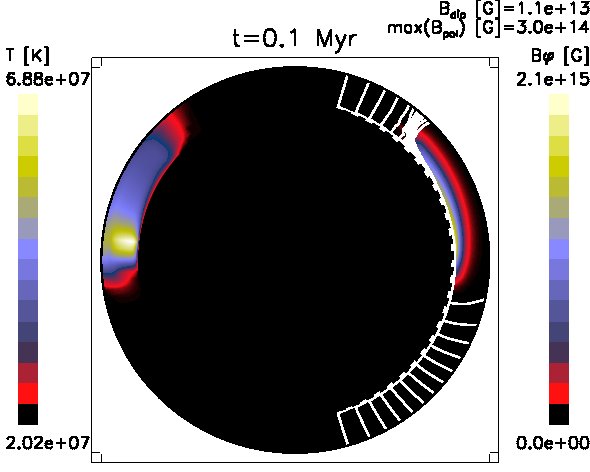}
\includegraphics[width=5.cm]{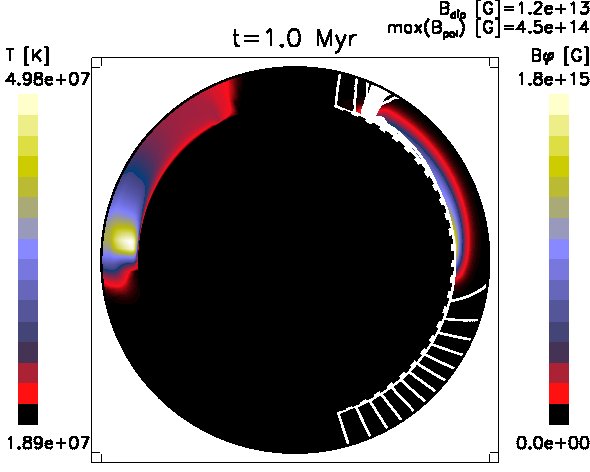}
\includegraphics[width=5.cm]{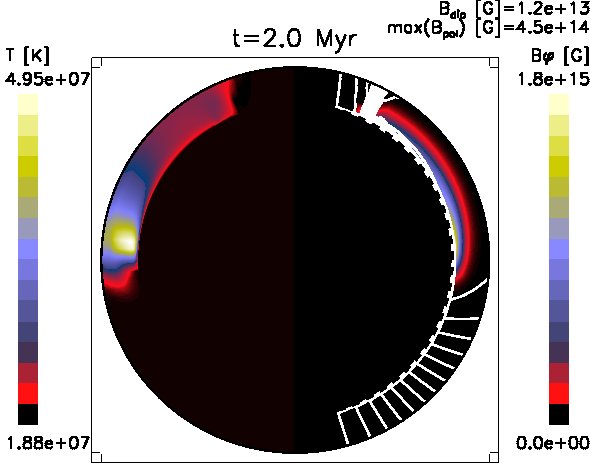}
\caption{Evolution for models AL, QL, BL, and BH (from top to bottom, respectively) at 100 kyr (left), 1 Myr (middle), and 2 Myr (right). The right hemispheres have the same meaning as Fig.~\ref{fig:initial_b}, while the left hemispheres represent the internal temperature. Again, the crust has been enlarged by a factor 4 and the lines penetrating in the core are not shown. The complete evolution of the crustal magnetic field up to $10$ Myr of models AL and BL is presented at \url{ftp://RY_URME_public:UR-Bremen1@ftp.dlr.de/download/MNRAS}.}
\label{fig:evo}
\end{figure*}

Such large values of $B$ do not affect the stability of the NS, since the magnetic pressure at these densities is about three orders of magnitude smaller than the hydrostatic pressure. On the other hand, such strong magnetic fields are expected to provoke strong stresses on the crust during the evolution, likely causing outburst activity, especially during the first millennia of life in the strongly magnetized NSs \citep{PP11}.

The other parameter we explore is the NS mass. Low-mass models, AL, QL and BL, have $M=1.4 M_{\odot}$, a radius of $11.6$ km and a crust thickness of $0.8$ km. Such NS structure is a representative for the so-called standard cooling models. In contrast, the high-mass model BH have $M=1.76 M_{\odot}$, a radius of $11.2$ km, and a crust thickness of only 0.5~km. Since the central density is larger than the low-mass stars, the neutrino cooling processes are faster due to the direct URCA channel \citep{PGW06}. Changing the mass of the NS has a direct effect on the magnetic field geometry, for a given $B_{\rm dip}$ and max($B_{\rm tor}$). As a matter of fact, a thinner crust implies a more curved magnetic field for the A configurations and different maximum values of $B$ (see Table~\ref{tab:models}).

\section{Results}\label{sec:results}

\subsection{Magnetic evolution}

\begin{figure}
   \centering
   \includegraphics[width=7.cm]{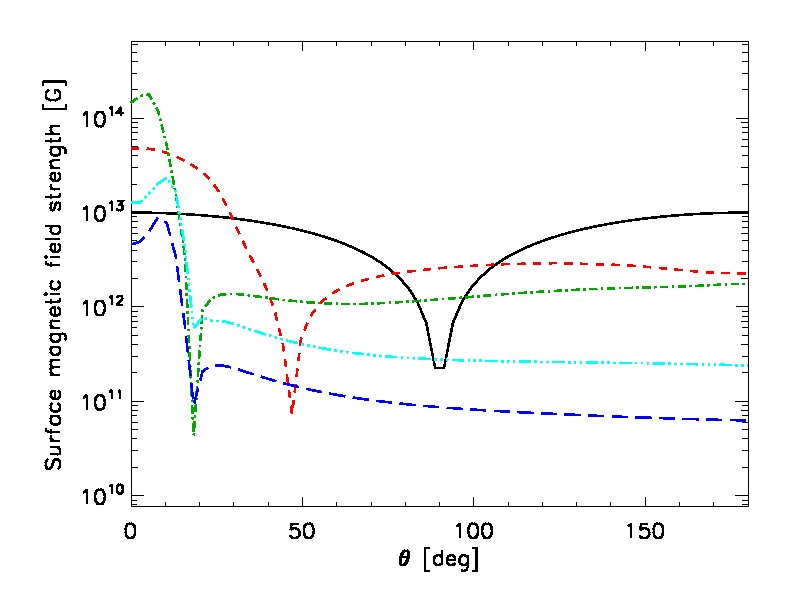}
   \includegraphics[width=7.cm]{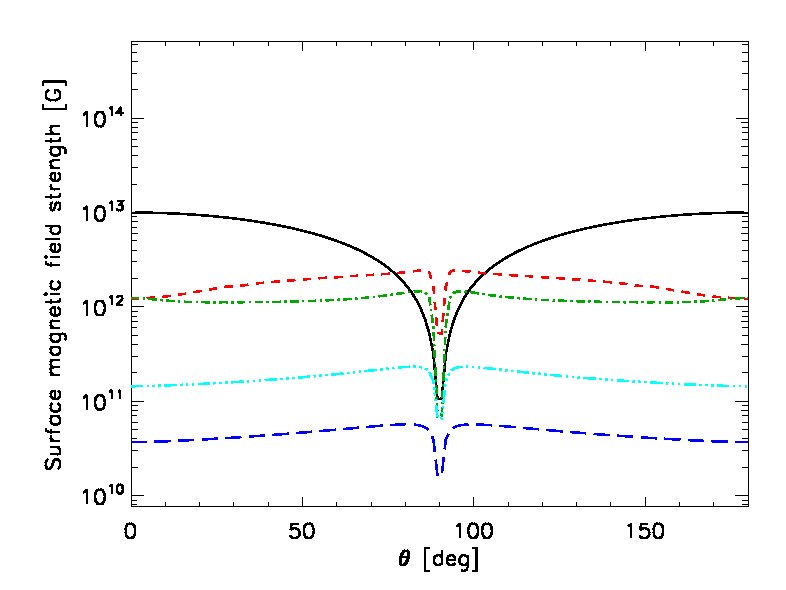}
   \includegraphics[width=7.cm]{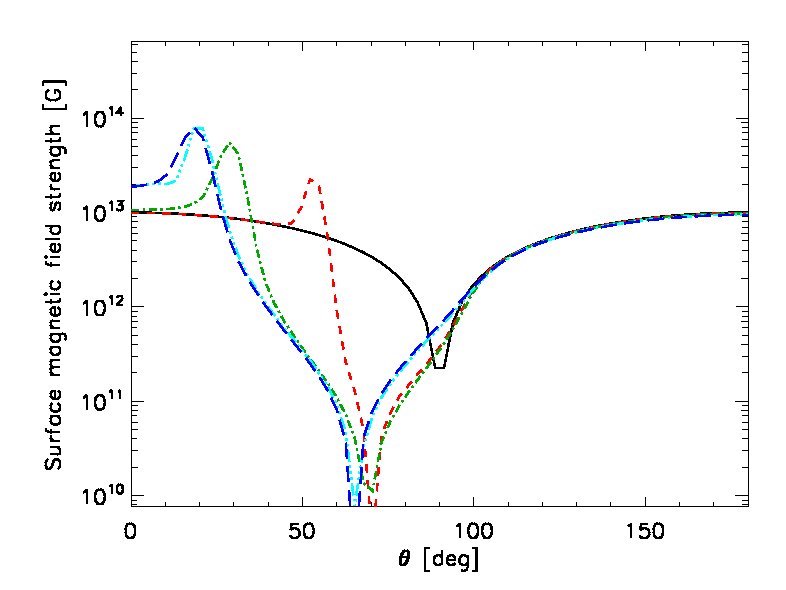}
   \includegraphics[width=7.cm]{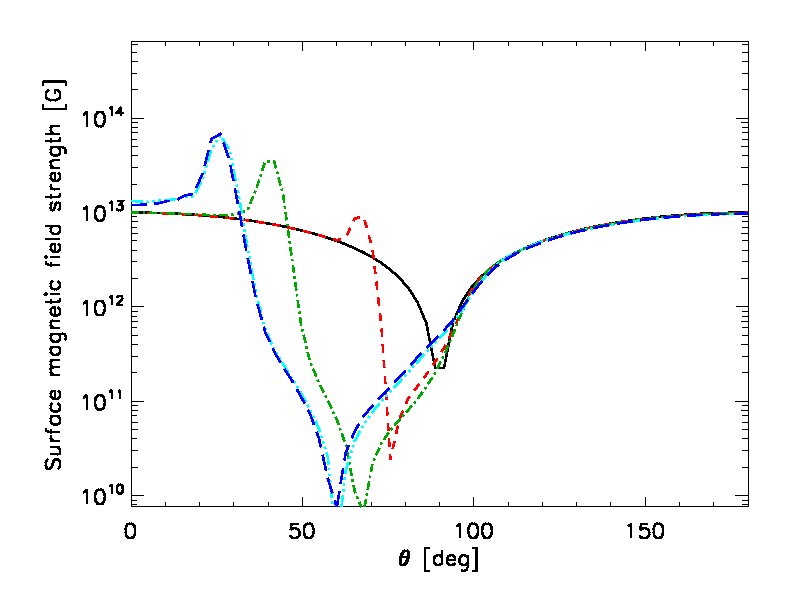}
 \caption{Evolution of magnetic field strength at $t=0$ (black solid line), 10 kyr (red dashes), 100 kyr (green dash-dotted), 1 Myr (cyan triple dot-dashed), and 2 Myr (blue long dashes) for models (from top to bottom): AL, QL, BL, and BH.}
\label{fig:bs}
\end{figure}

\begin{figure}
   \centering
   \includegraphics[width=7cm]{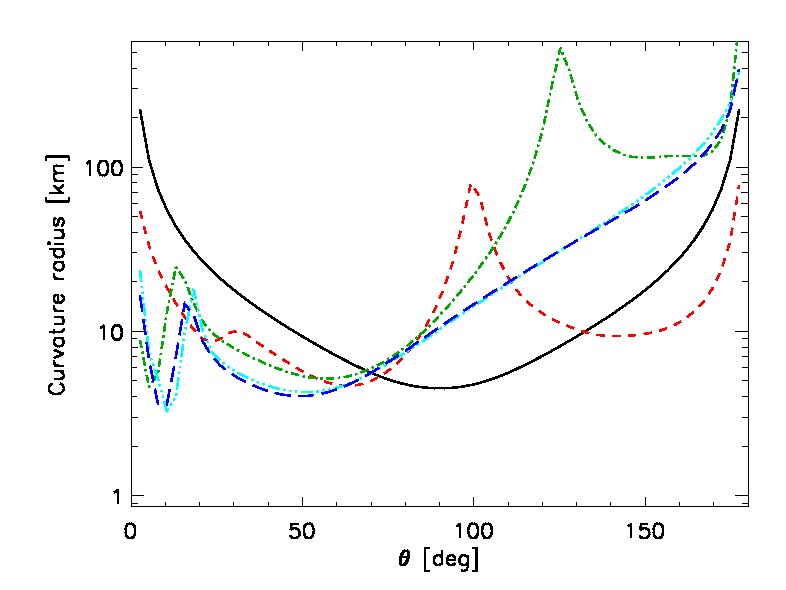}
   \includegraphics[width=7cm]{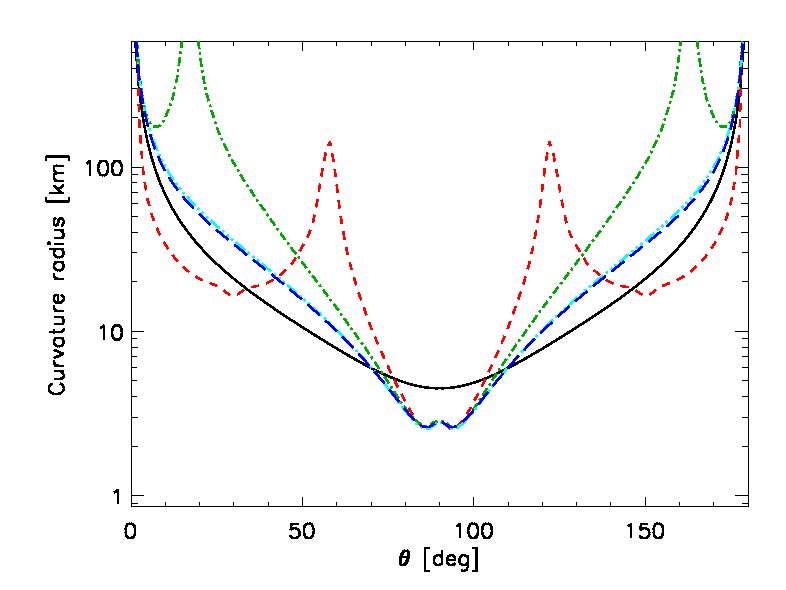}
   \includegraphics[width=7cm]{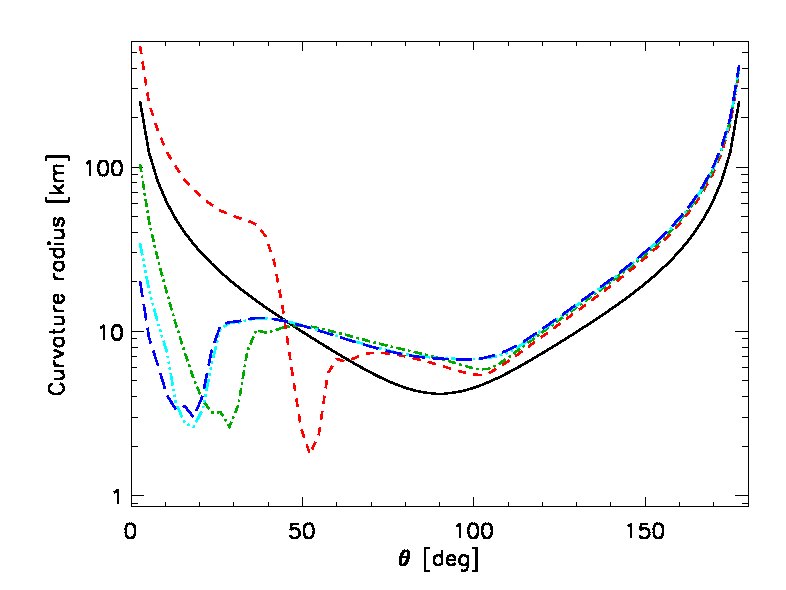}
   \includegraphics[width=7cm]{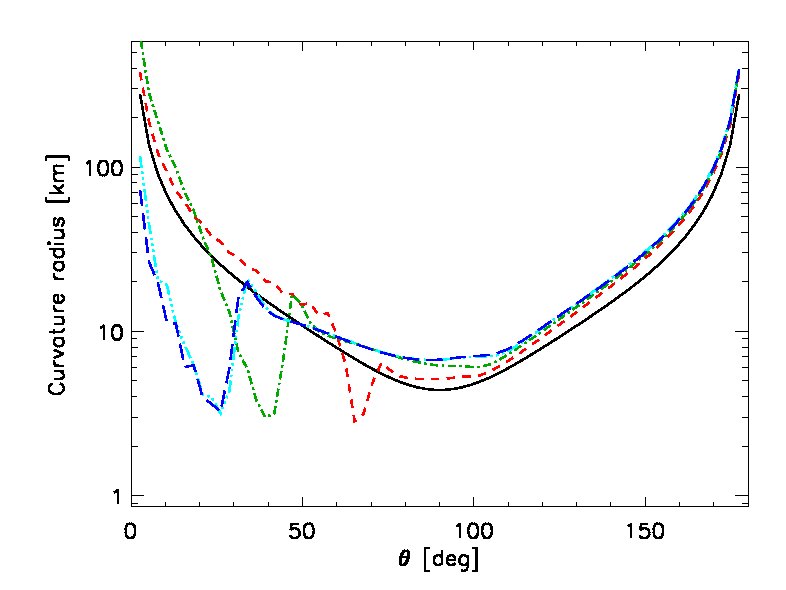}
 \caption{Evolution of the radius of curvature at $t=0$ (black solid line), 10 kyr (red dashes), 100 kyr (green dash-dotted), 1 Myr (cyan triple dot-dashed), and 2 Myr (blue long dashes) for models (from top to bottom): AL, QL, BL, and BH.}
\label{fig:rcurv}
\end{figure}

In Figs. \ref{fig:evo} we show the snapshots of the crustal magnetic field evolution, for the models of Table~\ref{tab:models}, respectively, after $10^5, 10^6$ and $2\times 10^6$ yr.\footnote{Movies which visualize the evolution of the crustal and of the external magnetic fields for the models that are presented here can be downloaded at \url{ftp://RY_URME_public:UR-Bremen1@ftp.dlr.de/download/MNRAS}}
Let us begin the discussion with the model AL (top panels). The very large toroidal field causes a drift of the currents towards the North pole in timescales of $\sim 10^4$ years. The drift is caused by the Burgers-like and advective terms in the Hall induction equation (see \cite{VCO00}, Appendix of \citealt{PG07} and eq.~(3.54) of \citealt{V13} for a deep analysis of the character of the Hall induction equation). Since the initial toroidal field is dipolar, the initital equatorial symmetry (see Fig.~\ref{fig:initial_b}) is soon broken, due to the preferred directionality of the drift. As a consequence, the currents and the magnetic field are advected towards the North pole. The toroidal field is strong enough to drag the much weaker poloidal field lines and compress them close to the pole. At small latitudes, the advective terms are weaker, and the drift stops, leading to the formation of a bunch of magnetic field lines at a latitude of $\sim 10^\circ$. In the region supporting the magnetic spot, where the magnetic field is more intense, the magnetic Reynolds number (see \cite{VPM12} for the definition and further discussion) reaches values up to $\sim 10^3$.

Meanwhile, the resistivity of the innermost part of the crust dissipates the local currents, especially where they are very strong, i.e. in the magnetic spot. The total magnetic energy in the crust is gradually dissipated, and both the dipolar magnetic field, $B_{\rm dip}$ and the maximum values of $B_{\rm pol}$ and $B_{\rm tor}$ (all indicated in each panel), also diminish. However, a sort of steady state is reached: the geometric configuration, consisting of a compact bunch of lines, is approximately maintained during millions of years. The reason for the relative local stability of the magnetic spot close to the North pole is the stable transfer of toroidal field energy into poloidal components and the settlement of the azimuthal currents at a certain depth. These currents maintain the small scale poloidal field structure and circulate preferentially at a radius where the electric conductivity has a local maximum, just outside the region with large $Q$, at $\rho \gtrsim 5\times 10^{13}$ g cm$^{-3}$.

The QL model evolves in a qualitatively different way. The even parity of the quadrupolar  toroidal field (multipolarity $l=2$) does not break the equatorial symmetry of the poloidal field component. Instead,  it creates a strong equatorial discontinuity, as in the models shown in \cite{VRPPAM13}. In this region, magnetic field lines are compressed. If the opposite parity had been chosen, then the drift would have led to a formation of two different bunches of lines towards the poles.

The most evident difference between models AL and BL (third row) is the slower evolution of the latter: the drift takes about $5\times 10^5$ years for the hot spot to find its final position in a meridional distance of about $30^{\circ}$ from the North pole. This is caused by the anchoring of the poloidal magnetic field to the highly conducting core (the $\theta$ and toroidal component of the magnetic field are discontinuous across the boundary, thanks to a strong current circulating in the outer core). Despite of this, the very strong toroidal field is able to drag the magnetic field lines towards the North, and creates a bunch of lines at $\sim 20^\circ$, some farther from the North pole compared with the model AL. This magnetic spot survives for a significantly longer period of time. Model BH (bottom panels) shows an evolution very similar to model BL. The main difference is that the magnetic spot is located farther from the pole. In general, the precise location of the bunch depends on the relative strength of the toroidal field and on the thickness of the crust.

During the evolution, in models AL and QL the ratio between the magnetic energy stored in the toroidal field, and the total magnetic energy, start from around $\sim 99.3\%$, then decreases down to $92\%$ and $95\%$, respectively, at $t\sim 1.2\times 10^5$ yr. After that, the ratio increases again, approaching unity. In model BL and BH, instead, the ratio is kept very close to 1 always, with most of the energy stored in the almost frozen crust. The on the long run energetically dominating toroidal field is in contradiction to the finding of \cite{KK12}, who see the poloidal field outweigh after the Hall period. However, they start with initial configurations where the toroidal energy is smaller than the poloidal one. 

The temporal evolution of the surface intensity of magnetic field, $B_s$, as a function of the meridional angle $\theta$, is presented in Fig. \ref{fig:bs}. In all models shown here, the local values of $B_s$ are high enough to support the PSG scenario ($B_s\gtrsim 5\times 10^{13}$ G) during a relatively large amount of time. The latter, in F of model AL, is less than 1 Myr, while, in case of models BL and BH, the lifetime of the magnetic spot is of at least several millions of years. Since the Ohmic decay has gradually dissipated the magnetic energy, the local $B_s$ decreases. In the QL model, the surface magnetic field in the spot is not high enough to ensure the required pair production.

An important parameter characterizing the magnetic spot is the radius of curvature of its lines, defined by $R_{\rm curv}\equiv ||-(\vec{b}\cdot\nabla)\vec{b}||^{-1}$, where $\vec b=\vec B/B$. We calculate $R_{\rm curv}$ just above the surface by supposing that the magnetospheric configuration is potential, i.e., a general superposition of vacuum multipoles (see how to apply such boundary condition in, e.g., Appendix A.5 of \citealt{V13}). Fig.~\ref{fig:rcurv} shows that in the magnetic spot, $R_{\rm curv}$ is much smaller than the one given by a dipolar field. Close to the North pole is $R_{\rm curv} \lesssim 10$ km, sufficiently small to enable the required pair creation cascade. Again, for QL the radius of curvature is below 10 km only close to the equator, only slightly less than for a pure dipole.

We found that these results are relatively insensitive to the initial poloidal field strength, as long as the magnetic field energy stored in it is small enough. Quite similar magnetic spots are created for $B_{\rm dip}\sim 5\times 10^{12}- 5\times 10^{13}$ G. Significantly stronger initial values of $B_{\rm dip}$ do not allow the formation of the bunch of lines because the lines are "strong" enough to oppose to the drift towards the pole, which is driven by the toroidal field.
We have also played with larger initial values of the crustal toroidal field. A maximum toroidal field strength there larger than $6\times 10^{15}$ G for model AL does not lead to the formation of a stable magnetic spot. In this case the Joule heat is enhanced and the magnetic field dissipation makes the magnetic spot disappear after about $10^5$ yrs.
The dipolar character and the strength of the toroidal field is a fundamental ingredient to reach these configurations. If, for instance, a quadrupolar toroidal field is chosen, then the drift will move currents and magnetic fields towards the equator \citep{VRPPAM13}, or towards both poles, depending on the relative polarity.

Last, as a test, we have run the same model as AL, but with low $Q=0.1$ everywhere, i.e., much larger values of conductivity. In this case, the Ohmic dissipation is strongly reduced, and the evolution is dominated by the Hall term. As a consequence, the toroidal field changes its orientation and location in a quasi-periodically way, pushing the magnetic spot from the northern to the southern hemisphere, as already noted by e.g. \cite{PG07} and \cite{PVG12}. As a consequence, the steady state of a bunch of lines is not achieved. Only during the first $10^5$ years a sufficiently strong magnetic spot close to the North pole appears. After about $2\times 10^6$ yr, more than one magnetic spot exist both at the northern and the southern hemisphere. However, each of them has a relatively small $B_s$. Thus, the existence of a highly resistive layer in deep crustal regions, $Q_{\rm ic}\gtrsim 10$, already supported by the clustering of periods of X-ray pulsars \citep{PVR13}, also favors the radio mechanism in the PSG model.

\subsection{Thermal evolution}

\begin{figure}
   \centering
   \includegraphics[width=7cm]{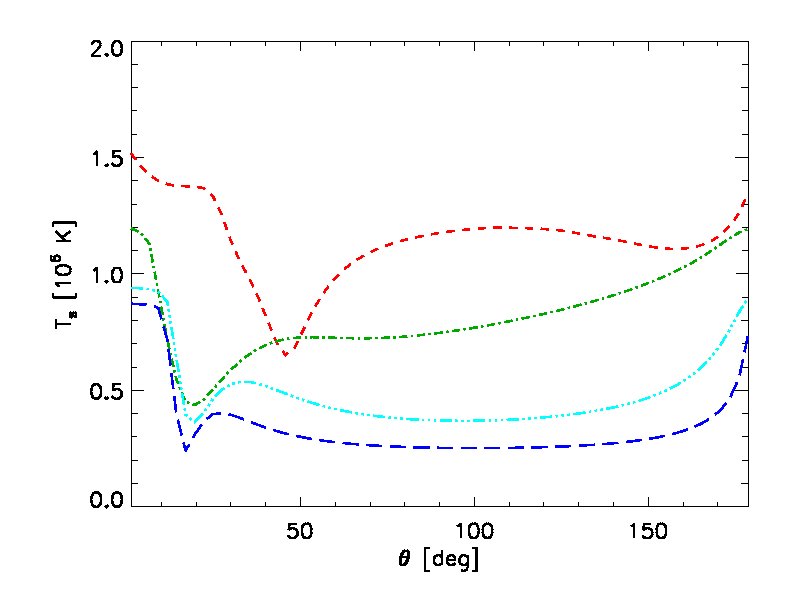}
   \includegraphics[width=7cm]{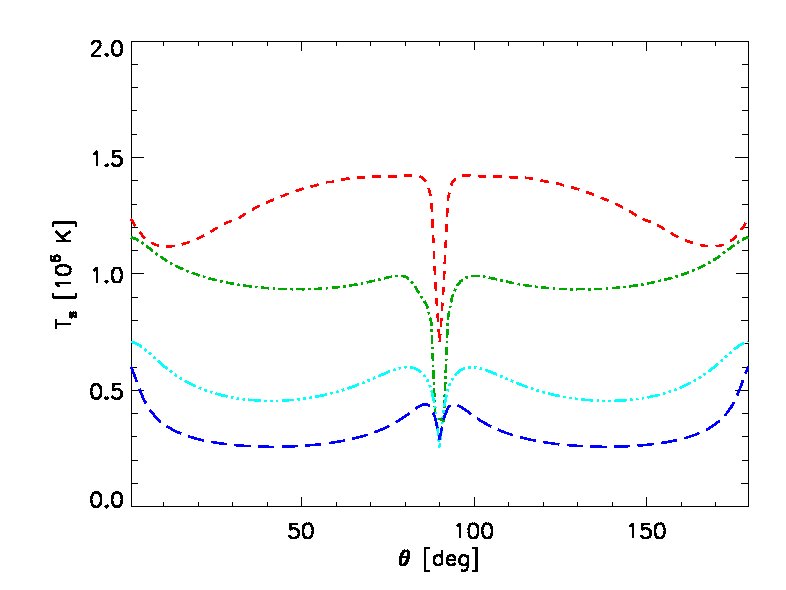}
   \includegraphics[width=7cm]{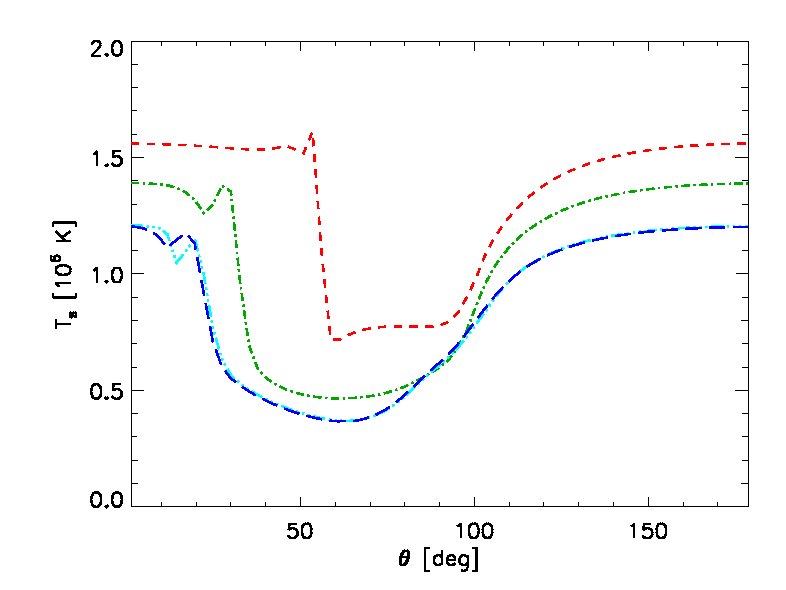}
   \includegraphics[width=7cm]{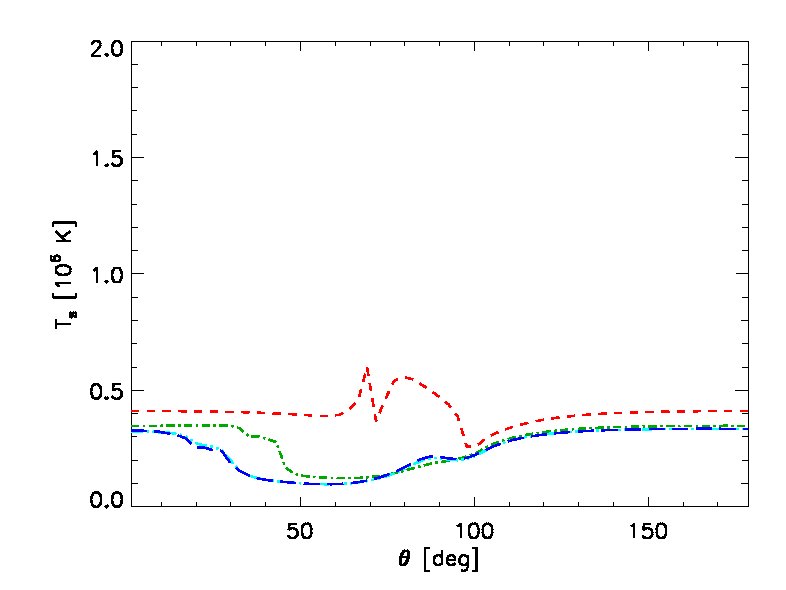}
 \caption{Evolution of the surface redshifted temperature at $t=10$ kyr (red dashes), 100 kyr (green dash-dotted), 1 Myr (cyan triple dot-dashed), and 2 Myr (blue long dashes) for models (from top to bottom): AL, QL, BL, and BH.}
\label{fig:ts}
\end{figure}

In the left hemispheres of Fig.~\ref{fig:evo} we show the evolution of the internal temperature. After $10^5$ yr (left panels), our models show the Northern hemisphere hotter than the rest of the star, due to the strong Joule heating associated with the bunch of magnetic field lines, and the toroidal field mostly extended in that hemisphere. The insulating effects of the tangential lines, and the absence of strong Joule heating in the South hemisphere make the latter colder than the core. The poles, where the magnetic field lines are radial, and no currents circulate, are always thermally connected with the core.

In model QL, a lot of heat is deposited in the equatorial regions, which are hotter than the poles and the core at the beginning. At later times, most of the Joule heating diffuse to the highly conductive core, and the differences inside the star are much less than at the beginning. In models AL and QL, such differences are almost inexistent after 1 Myr, due to the faster dissipation of the magnetic energy, which decreases the anisotropy in the conductivity.

In models BL and BH, in which the toroidal field is larger, and the dissipation is slower due to the geometry of the configuration, the differences in $T$ caused by the inhomogeneous Joule heating are important even after millions of  years. Regions at middle latitudes, connected with the bulk of the toroidal field, are hotter than the rest, and insulated from the core. The internal temperature distribution is maintained approximately similar, due to the reach of the steady state: the Joule heating continuously compensates the cooling. A common feature is that the outer crust in the region of the bunch of lines is slightly colder than the rest at late times, due to the insulating effects.

A striking difference is seen between models BH and BL. Since the NS of models BH is a massive one, it is subject to accelerated cooling by enhanced neutrino emission in DURCA processes in the core. The latter, actually, is colder than the crust, since it acts as a heat sink. The internal temperatures for this models are almost one order of magnitude lower than in BL. The presence of Joule heating can partially or totally erase such difference, keeping the surface hot, if the currents are strong enough and located in a region close enough to the surface (e.g., \citealt{KKPY14}).

The redshifted surface temperature $T_s(\theta)$ is shown in Fig. \ref{fig:ts}, for the same times as before (except the black solid line, $t=0$, which would reflect only the arbitrarily chosen initial temperature). The differences in surface gravity between models AL/QL/BL (low mass) and BH (high mass) give differences of $\sim 10\%$ to their $T_s(\theta)$, for the same physical temperature. The surface temperature depends on the envelope model, which takes into account the temperature at the crust/envelope interface, $T_b$, and the local magnetic field intensity and direction. The envelope hosts the maximum gradients of temperature, pressure and density, and can provide great differences between the temperature profiles $T_b(\theta)$ and $T_s(\theta)$. Such blanketing effect is very effective in the regions with strong tangential magnetic field, and it is evident if one compares Fig.~\ref{fig:ts} and Fig.~\ref{fig:evo}.

Looking at model AL (top panel), we see that $T_s$ reflects the magnetic drifting toward the North pole during the first $\sim 10$ kyr (black and red lines). At later times, when the magnetic spot is settled, a small hot spot is created (green, cyan and blue lines). Note that the $T_s$ profile does not reflect the crustal temperature, which has a maximum close to the magnetic spot, due to the strong local Joule heating. This is due to the strong insulating effect of the envelope where the poloidal field lines are almost parallel to the surface. The insulating effect leads to a steep decrease of the surface temperature going south, until a minimum is reached at $\sim 20^\circ$ from the North pole. The temperature there is $\sim 35\%$ of $T_s$ at the North pole. Southward, $T_s$ increases again due to the change in the directionality of the poloidal field lines, which partially allow the transport of heat coming from the deep crustal layers. Throughout the evolution, the poloidal field lines are almost radial very close to the poles: the heat coming from the core easily flow to the surface, making the poles quite hot.

In model QL (second panel), initially the equatorial region is the hottest part of the surface, due tot he very large interior temperature in those regions. Later, the blanketing effect reduces such differences, and the surface does not present large inhomogeneities in $T_s$.

The case of BL (third panel) is qualitatively similar. The differences are that the North pole is hotter than in model AL, and the minimum is very broad. This is due to the very large region with almost tangential magnetic field lines (see the second row of Fig.~\ref{fig:evo}). Model BH (bottom panel), on the other hand, show much lower temperatures, due to the colder core. The differences in temperature between the hot and cold region are more limited, and the profile is smoother.

Note that the radiation emitted from regions with surface temperatures $T_s\gtrsim 10^6$ K are potentially observable by X-ray telescopes. In this sense, models AL and BL would correspond to X-ray detectable NSs, provided that their interstellar absorption is not too large, and the viewing angle is favourable. If detected, the spectral fits would provide relatively small emitting area \citep{PVPR13}. On the other hand, models like BH are colder, which makes the surface condensation easier.

\section{Conclusions}\label{sec:conclusions}

In this work we have shown simulations of the magnetic and thermal evolution which support the scenario of the PSG, which require the presence of at least one long-living ($\gtrsim 1$ Myr) magnetic spot, i.e., a region of the surface with a strong and curved magnetic field. The initial possible magnetic configurations are virtually infinite, and poorly constrained by theoretical and observational studies.

Before coming to the main conclusions, there are a number of caveats we should remind the reader, about some intrinsic limitations of our magneto-thermal simulations run with the code of \cite{VPM12}. First, the initial configurations are likely simplifying the reality, ignoring more complicated geometries (e.g., strong multipolar components). Second, simulations are 2D: the creation, size and timescales of magnetic spots could be different when azimuthal dependences are allowed. Third, our calculations do not consider processes that may lead to the interchange of magnetic energy between the core and the crust, and between the crust and the magnetosphere. The important MHD processes in the core are still unclear, and could include buoyancy, ambipolar diffusion and interactions between the superfluid vortices and the superconducting fluxoids. In our simulations, the core has very little influence on the global evolution, because the magnetic field evolution therein is driven by the very slow Ohmic dissipation (models BL and BH), or the core is treated as a type-II superconductor (models AL and QL). However, the core contains most of the volume of the NS, and could have an important imprint: how and how fast can the magnetic field migrate from the core to the crust?

In this sense, the analytical study and the inclusion of such effects in the simulations, and/or an upgrade to 3D could provide different quantitative results, in terms of values given above. Similarly, the critical magnetic field needed to sustain the gap (by means of a strong cohesive energy of the magnetized surface) is a fundamental ingredient in the PSG model, but it is far from being well known, relying solely on the pioneering work of \cite{ML07}, and being strongly dependent on the temperature. All these factors should be kept in mind when interpreting the quantitative numbers.

On the other hand, we have found a number of conditions that seem in general necessary for the creation and maintenance of the magnetic spots:

\begin{enumerate}
\item lower limits for the initial magnetic field strength: $B_{\rm dip} \gtrsim 5\times 10^{12}$ G for the polar surface dipolar field and max$(B_{\rm tor}^{\rm cr})\gtrsim 10^{15}$ G for the maximum toroidal field in the crust;
\item upper limits for the initial magnetic field strength: $B_{\rm dip} \lesssim 5\times 10^{13}$ G for the polar surface dipolar field and max$(B_{\rm tor}^{\rm cr})\lesssim 6\times 10^{15}$ G for the maximum toroidal field in the crust;
\item the initial toroidal field has to have a odd multi-polar structure, preferentially a dipolar one; quadrupolar or initial fields of higher multipolarity provide magnetic spots too, but their field strength is too low, and their radius of curvature is too large, to guarantee the PSG conditions;
\item in the deeper regions of the crust the electric conductivity must be relatively low, compatible with an impurity parameter $Q_{\rm ic}\sim 10-100$.
\end{enumerate}
One could ask whether these conditions are fulfilled in a large fraction of newborn NSs, and if this fraction is compatible with the observed population of radio pulsars. Note that (i) and (ii) mean that almost $100\%$ of the initial magnetic energy must be stored in its toroidal component (see Table~\ref{tab:models}). The creation of such very strong toroidal fields is not easily justified, and a possible explanation relies on the first $\sim 20$ seconds of a NS life, when the strong differential rotation could wind up an initially relatively weak poloidal field component to an extremely strong toroidal one \citep{P79,S02}.

The limits on $B_{\rm dip}$ are particularly interesting if one looks at the radio-quiet population of NSs (seen in X-rays), since the latter are often highly magnetized \citep{RPTT12}. Another important conclusion is that magnetic spots are being created both if the initial magnetic field is MHD equilibrium-like (models B), and if the toroidal field is extended throughout the crust (model A), respectively. The lifetime of the magnetic spot in case of model B is even significantly longer than in model A. Since the currents and the magnetic field of models B live longer, the local surface field strength in the spot is after $\sim 10^{7}$ years still in the order of $10^{14}$ G, much larger than in case of model A.

From an observable point of view, one could require the lifetime of the magnetic spots to be similar to the radio pulsars ages. The latter are usually estimated by means of the spin-down age, which spans a large range, $\tau=P/2\dot{P}\sim 10^3-10^8$ yr. However, the latter strongly overestimates the real age if the dipolar field $B_{\rm dip}$ decreases fast, like in model AL. As a consequence, on this ground none of the two scenarios can be ruled out, since we do not know neither how many pulsars we are missing, nor the real ages of the radio pulsars. Actually, the crustal confined models, like AL, are favored by the timing and spectral analysis of the X-ray thermally emitting NSs \citep{VRPPAM13}, and by the most recent population synthesis studies, which try to reproduce the observed $P$-$\dot{P}$ distribution taking into account observational and physical effects \citep{GMVP14}. A population synthesis that includes and compares both radio and X-ray data could also provide some further constraints on the models proposed here, if one considers their different surface temperatures, which imply different detectability, as discussed above.

An intermediate scenario is proposed by the MHD equilibrium configuration including a new treatment for the type-I superconducting core \citep{L14}: a magnetic field threading both the crust and the core, but with a concentration in the former, especially suitable for the less magnetized NSs ($B_{\rm dip}\lesssim 5\times 10^{13}$ G, like in our models). Such configuration, whose stability and unicity have not been proven, could, in principle, provide both the strong currents in the crust and long-lived magnetic spots.

A possible problem of our models is that the drift takes some time to create the magnetic spot, since we start from a large-scale dipole. The associated initially large-scale field lines are progressively bent to create the required small-scale field. Before that, i.e. during the first several $10^4$ yr, the surface temperature is quite large and the the radius of curvature is large. Therefore, how can young NSs, like Crab and Vela pulsars, produce radio emission in the PSG scenario? We note that the initial formation time is totally related to the choice of an initial dipole for the poloidal magnetic field. Thus, a possible explanation is that the initial magnetic field has, since the beginning, multipolar structure in the poloidal field, which can soon provide the small scales.

We have also shown that the conditions for the magnetic spot weakly depend on the NS mass: they appear and evolve similarly in both standard coolers (low masses, models AL and BL) and in the fast coolers (BH), even if the temperature for the two kinds of models is very different. This implies that other changes in the temperature evolution caused by a change of, e.g., the superfluid gap or the envelope model, will not affect the appearence and dynamics of the magnetic spots.

In a short conclusion, we confirm that the Hall drift, combined with a huge reservoir of toroidal magnetic energy stored deep in the crust, can provide the required structures on a correct time scale.

\section*{Acknowledgments}
UG is partially financed by the Grant DEC-2012/05/B/ST9/03924 of the Polish National Science Center. DV is supported by grants AYA2012-39303 and SGR2009-811. We acknowledge the many enlightening discussions with Jos\'{e} Pons. We are also indebted to Janusz Gil, Gogi Melikidze and Andrzej Szary for numerous fruitful discussions and encouragements. Last but not least we are grateful to an anonymous referee whose questions, comments and suggestions improved the manuscript significantly.

\bibliography{pulsars}

\label{lastpage}
\end{document}